\documentclass[final]{svjour3}
\usepackage{graphicx}
\usepackage{rotating}
\usepackage{amssymb}
\usepackage{mathptmx}
\usepackage[numbers, compress ]{natbib}
\usepackage{hyperref}

\usepackage{xcolor}
\usepackage{textgreek}
\usepackage[rightcaption]{sidecap}
\sidecaptionvpos{figure}{c}

\makeatletter
\journalname{Journal of Low Temperature Physics}

\bibpunct{}{}{,}{s}{}{,}

\begin{document}

\newcommand{\hdblarrow}{H\makebox[0.9ex][l]{$\downdownarrows$}-}
\title{Detector Array Readout with Traveling Wave Amplifiers}

\author{A.~Giachero$^{1,2}$ \and
        C.~Barone$^{3,4}$\and
        M.~Borghesi$^{1,2}$\and
        G.~Carapella$^{3,4}$\and
        A.P.~Caricato$^{5,6}$\and
        I.~Carusotto$^{7,8,9}$\and
        W.~Chang$^{19}$\and
        A.~Cian$^{10,9}$\and
        D.~Di~Gioacchino$^{11}$\and
        E.~Enrico$^{12,9}$\and
        P.~Falferi$^{10,13,9}$\and
        L.~Fasolo$^{14,12}$\and
        M.~Faverzani$^{1,2}$\and
        E.~Ferri$^{1,2}$\and
        G.~Filatrella$^{15,4}$\and
        C.~Gatti$^{11}$\and
        D.~Giubertoni$^{10,9}$\and
        A.~Greco$^{14,12}$\and
        C.~Kutlu$^{20,19}$\and
        A.~Leo$^{5,6}$\and
        C.~Ligi$^{11}$,
        G.~Maccarrone$^{11}$\and
        B.~Margesin$^{10,9}$\and
        G.~Maruccio$^{5,6}$\and
        A.~Matlashov$^{19}$\and
        C.~Mauro$^{4}$\and
        R.~Mezzena$^{16,9}$\and
        A.G.~Monteduro$^{5,6}$\and
        A.~Nucciotti$^{1,2}$\and
        L.~Oberto$^{12,9}$\and
        S.~Pagano$^{3,4}$\and
        V.~Pierro$^{17,4}$\and
        L.~Piersanti$^{11}$\and
        M.~Rajteri$^{12,18}$\and
        S.~Rizzato$^{5,6}$\and
        Y.K.~Semertzidis$^{19,20}$\and
        S.~Uchaikin$^{19}$\and
        A.~Vinante$^{13,10,9}$
}


\institute{
  $^1$University of Milano Bicocca, Department of Physics, I-20126 Milan, Italy\\
  $^2$INFN - Milano Bicocca, I-20126 Milan, Italy\\ 
  $^3$University of Salerno, Department of Physics, I-84084 Fisciano, Salerno, Italy\\
  $^4$INFN - Napoli, Salerno group, I-84084 Fisciano, Salerno, Italy\\
  $^5$University of Salento, Department of Physics, I-73100 Lecce, Italy\\
  $^6$INFN – Sezione di Lecce, I-73100 Lecce, Italy\\
  $^7$INO-CNR BEC Center, I-38123 Povo, Trento, Italy\\
  $^8$University of Trento, Department of Physics, I-38123, Povo, Trento, Italy\\
  $^9$INFN - Trento Institute for Fundamental Physics and Applications, I-38123, Povo, Trento, Italy\\  
  $^{10}$Fondazione Bruno Kessler, I-38123, Povo, Trento, Italy\\
  $^{11}$INFN - Laboratori Nazionali di Frascati, I-00044, Frascati, Rome, Italy\\
  $^{12}$INRiM - Istituto Nazionale di Ricerca Metrologica, I-10135 Turin, Italy\\
  $^{13}$IFN-CNR, I-38123 Povo, Trento, Italy\\
  $^{14}$Polytechnic University of Turin, I-10129 Turin, Italy\\
  $^{15}$University of Sannio, Department of Science and Technology, I-82100, Benevento\\
  $^{16}$University of Trento, Department of Physics, I-38123, Povo, Trento, Italy\\
  $^{17}$University of Sannio, Department of Engineering, I-82100 Benevento, Italy\\
  $^{18}$INFN - Torino,  I-10125 Turin, Italy\\
  $^{19}$Center for Axion and Precision Physics Research, Institute for Basic Science (IBS), Daejeon 34051, Republic of Korea\\
  $^{20}$Department of Physics, Korea Advanced Institute of Science and Technology (KAIST), Daejeon 34141, Republic of Korea\\ 
  \email{andrea.giachero@mib.infn.it}\\
}

\maketitle

\begin{abstract}
Noise at the quantum limit over a large bandwidth is a fundamental requirement for future applications operating at millikelvin temperatures, such as the neutrino mass measurement, the next-generation x-ray observatory, the CMB measurement, the dark matter and axion detection, and the rapid high-fidelity readout of superconducting qubits. The read out sensitivity of arrays of microcalorimeter detectors, resonant axion-detectors, and qubits, is currently limited by the noise temperature and bandwidth of the cryogenic amplifiers. The DARTWARS (Detector Array Readout with Traveling Wave AmplifieRS) project has the goal of developing high-performing innovative traveling wave parametric amplifiers (TWPAs) with a high gain, a high saturation power, and a quantum-limited or nearly quantum-limited noise. The practical development follows two different promising approaches, one based on the Josephson junctions and the other one based on the kinetic inductance of a high-resistivity superconductor. 
In this contribution we present the aims of the project, the adopted design solutions and preliminary results from simulations and measurements.


\end{abstract}

\section{Introduction}


A wide range of quantum technologies relies on the faithful detection of microwave signals at millikelvin temperatures. This includes quantum computing with superconducting~\cite{Krantz2019} and spin~\cite{Stehlik2015} qubits, multiplexed readout of particle and astronomical detectors (TESs~\cite{Irwin-Hilton2005,Gottardi2021} and MKIDs~\cite{Day2003,Zmuidzinas2012}) or the search for axionic dark matter with microwaves cavities~\cite{Alesini2019,Crescini2020}. These experiments often use amplifiers based on semiconductors (HEMT) with an high gain but with a noise 10–40 times above the standard quantum limit, the fundamental limit imposed by quantum mechanics~\cite{Caves1982}. 
At microwave frequencies, Josephson Parametric Amplifiers (JPA)~\cite{Castellanos2007} have demonstrated quantum-limited noise and are currently used in axion search and qubit read out. 
JPAs are still insufficient for addressing the needs of the above-mentioned applications. An alternative and innovative solution is based on parametric amplification exploiting the traveling wave concept. A Traveling Wave Parametric Amplifiers (TWPA) is designed by exploiting the signal (nonlinear) response of reactive parts (typically inductance) in a superconducting circuit. A large pump tone modulates this inductance, coupling the pump ($f_p$) to a signal ($f_s$) and idler ($f_i$) tone via frequency mixing such that $2f_p=f_s+f_i$ (4-wave mixing, 4WM) or $f_p=f_s+f_i$ (3-wave mixing, 3WM). The nonlinear inductance can be implemented by means of Josephson Junction (JJ) and Kinetic Inductance (KI) of superconductors. The relationship is, at the first order, $L(I)=L_0 [1+(I/I_c)^2]$, where in JJ, $I_c$ is the junction critical current, while in KI, $I_c$ is the superconductor critical current. At $I < I_c$, junctions are dissipationless and act as nonlinear inductors. 

In recent times, the concept of a parametric amplifier with microwaves travelling along a transmission line with embedded Josephson junctions (TWJPA) was developed in several groups~\cite{Brien2014,White2015}. These devices have already demonstrated quantum-limited noise and are used to readout superconducting qubit~\cite{Macklin2015}. 
TWPA based on KI, called Dispersion-engineered Traveling Wave Kinetic Inductance (DTWKI) parametric amplifier or Kinetic Inductance Traveling Wave Parametric amplifier (KI-TWPA or in short-form  KIT), was originally proposed by J.~Zmuidzinas group at Caltech in 2012 and the first demonstrators have been produced after few years of development by Caltech~\cite{Eom2012} and NIST~\cite{Vissers2016}. Recent results showed a noise very close to the quantum limited performance with a gain around 15 dB\,over a 5\,GHz bandwidth~\cite{Zobrist2019, Malnou2021}. These results highlight the potential of parametric amplifiers. However, they also show that significant progress is still needed to achieve the required gain, bandwidth, and noise performance with acceptable power dissipation.

The Detector Array Readout with Traveling Wave Amplifiers (DARTWARS) is a three-year project funded by the Italian Institute for Nuclear Physics (INFN) starting from 2021 that aims to fulfil these requirements. 
The technical goal is to achieve a gain value around 20\,dB, comparable to HEMT, a high saturation power (around -50\,dBm), and a quantum limited or nearly quantum limited noise ($T_N < 600$\,mK). These features will lead to reading out large arrays of detectors or qubits with no noise degradation. 
The development of two different approaches is surely more demanding in terms of time and costs but, for sure, is the best way to maximize the success of the action.

\section{TWJPA Design}
Josephson junctions based Traveling Wave Parametric Amplifiers are composed by a coplanar waveguide (CPW) embedding a chain elementary cells containing Josephson Junctions (eg. rf-SQUIDs). The nonlinear behaviour can be tuned using an external magnetic field or a DC bias that flows in the signal line, allowing it to work as a three-wave mixer~\cite{Zorin2016} or a four-wave mixer. The DARTWARS design follows the quantum model developed by INRiM~\cite{Greco2020,Fasolo2021}, a coupled mode equation approach to describe the behaviour of a TWJPA in the few-photons or even single-photon level.

\begin{figure}[!t] 
 \begin{center}
    \includegraphics[clip=true,width=\textwidth]{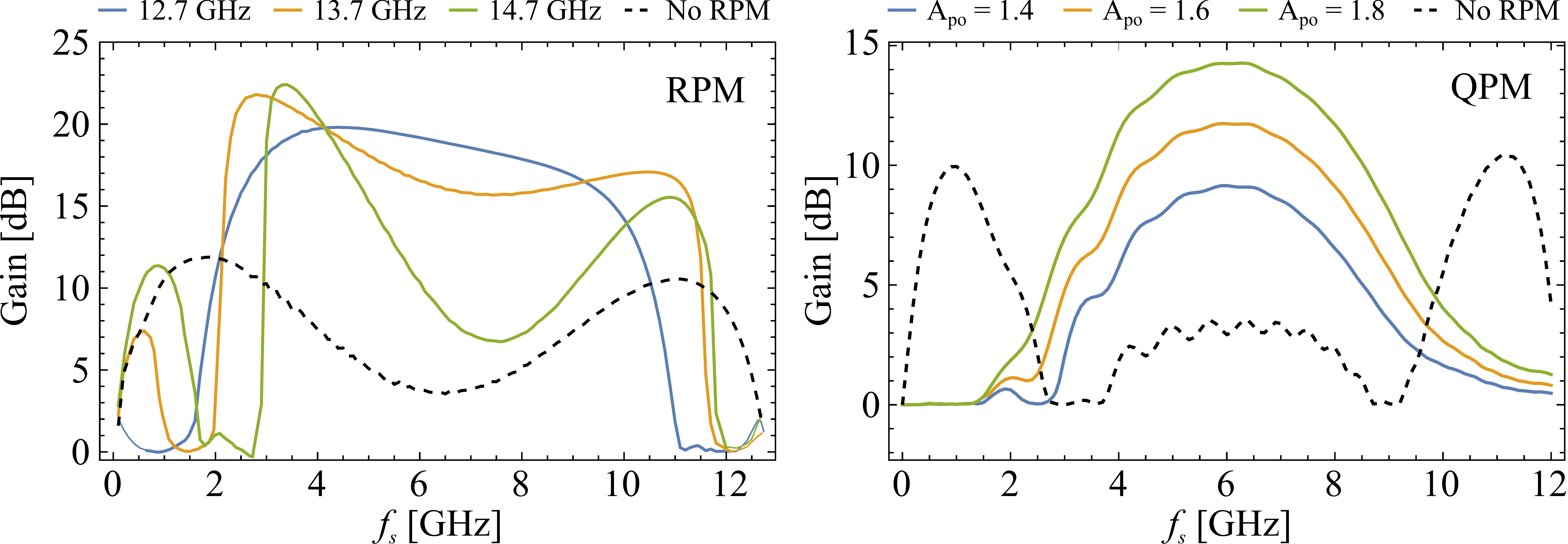}
  \end{center}
 \caption{\label{fig:RPMQPM} (left) Gain in the 3WM regime as a function of signal frequency $f_s$ for a JTWPA implementing the RPM scheme calculated for different pump frequencies by exploiting the Coupled Modes Equations approach (CME-2). 
 (right) Gain in the Three-Wave Mixing regime as a function of signal frequency $f_s$ for a JTWPA implementing the QPM scheme, calculated for different normalized pump amplitudes ($A_{po}$) by exploiting the Coupled Modes Equations approach (CME-2).
 }
\end{figure}

To avoid power leakage into higher frequency tones the CPWs are equipped with a modified dispersion relation following two different approaches, the Resonant Phase Matching (RPM) and the Quasi-Phase Matching (QPM) techniques. RPM (figure \ref{fig:RPMQPM}, left) uses a reduced plasma frequency mixed to a periodic load of LC resonators to create mismatch among the traveling tones, in order to suppress higher harmonic generation, and re-phrase just the signal tone that is meant to be amplified through the opening of a bandgap in the dispersion relation~\cite{OBrien2014}. On the contrary, QPM (Fig. \ref{fig:RPMQPM}, right) uses a mix of low plasma frequency and a sign modulation of the nonlinearity into the medium to suppress higher harmonic generation and stimulate amplification by changing the phase of the traveling waves of π after a coherence length has been reached~\cite{Zorin2021}.

During 2021 INRiM developed a modified version of the quantum model~\cite{Greco2020}, which implements in its mathematical structure the modified dispersion relation given by the RPM technique. The study highlights how the presence of the RPM helps reduce the phase mismatch between the traveling tones given by the low plasma frequency (35 GHz), enhancing the overall gain of several dB respect to the case where RPM is not used. Moreover, a numerical study on the QPM approach has been performed through a Coupled Mode Equations (CME) method~\cite{Dixon2020}. The CME method takes into account modes at frequency as high as the fifth pump harmonic, meaning that the power dispersion of the pump into higher harmonics can be quantitatively studied and monitored.

The physical implementation of the circuit components occurred through electromagnetic simulations performed with the software Sonnet Suite.
The RPM technique needs shunt resonators with resonant frequency very close to the pump frequency, hence in the range 10-15 GHz depending on the working band, and a reduced plasma frequency, around 35 GHz. These facts require the presence of large capacitances in the resonators and shunting the Josephson junctions, that can be reached by adding an extra layer of dielectric material in the layout so as to allow the use of parallel plate capacitors. The same result can be achieved using extended 2D planar resonators and large area Josephson junctions. The former approach allows reducing the areas of the Josephson junctions at the cost of adding a layer of dielectric material. 
The Sonnet Suite has been used to realise the physical layout of the devices in the presence and absence of the extra dielectric layer, that has been chosen to be SiO2 and a-Si:H.
The physical implementation of QPM-JTWPA does not require resonators but just a poling of the rf-SQUIDs and a reduced plasma frequency, that again can be reached by enlarging the area of the Josephson junctions.

\section{TWJPA Fabrication}
The preliminary characterization of the first TWJPA prototypes fabricated at INRiM highlighted the presence of unwanted effects ascribable to an inhomogeneity in the area of the fabricated Josephson Junctions (JJs). These first prototypes were fabricated exploiting a well-established nanofabrication technology based on electron beam lithography (EBL) on a double layer polymeric mask (bottom layer realized with the copolymer MMA(8.5)MAA EL11 and top layer realized with 950K PMMA A4, both by KayakuAM) followed by an Aluminium e-gun evaporation exploiting the Niemeyer-Dolan technique. The realization of thousands of micrometric structures, such as those present in a TWJPA architecture, in the range of time in which the lithographic system is stable and predictable requires the use of a large beam current, causing a reduction in the pattern resolution and hence the origin of a spread in the areas of the JJs. To overcome this issue, in the first half of 2021, INRiM started to fabricate new TWJPAs exploiting a two-step lithography approach. The lithography of the components with a resolution bigger than 1 µm were realized with an UV laser writer (µPG-101 from Heidelberg) exploiting a polymeric mask realized with the reverse image photoresist AZ 5214 E from Microchemicals GmbH. To reach the desired resolution for this process over large areas, up to 2" diameter, many process parameters have been tuned (spinning protocol, power of the UV laser, bake time and temperature of the mask and time of the development). The second lithography step, reserved to the features smaller than 1 µm, will be realized with EBL. This will guarantee the use of a small beam current and hence an increase in the resolution of the process. Moreover, to further reduce the JJ areas inhomogeneity due to the overlap of unpredictable rounded edges, a new design for these structures is under development. This design will exploit a new double layer mask made with LOR 20B and 950K PMMA A4, both from KayakuAM. Preliminary characterization of the lithography and development of this mask have been already performed during the first half of 2021. To realize TWJPA architectures that need the presence of a dielectric layer, during the first half of 2021 INRiM grown and characterized thin films of SiO2 and a-Si:H realized with an Inductively Coupled Plasma Chemical Vapour Deposition system (PlasmaPro 100 ICPCVD by Oxford Instruments). Several samples, realized at different substrate temperatures (from 200 to 350\,$^\circ$C), were preliminary analyzed with spectroscopic ellipsometric and FTIR (Fourier-transform infrared spectroscopy) measurement.

\section{KI-TWPA Design}
\begin{SCfigure}[][!t]
\includegraphics[clip=true,width=0.5\textwidth]{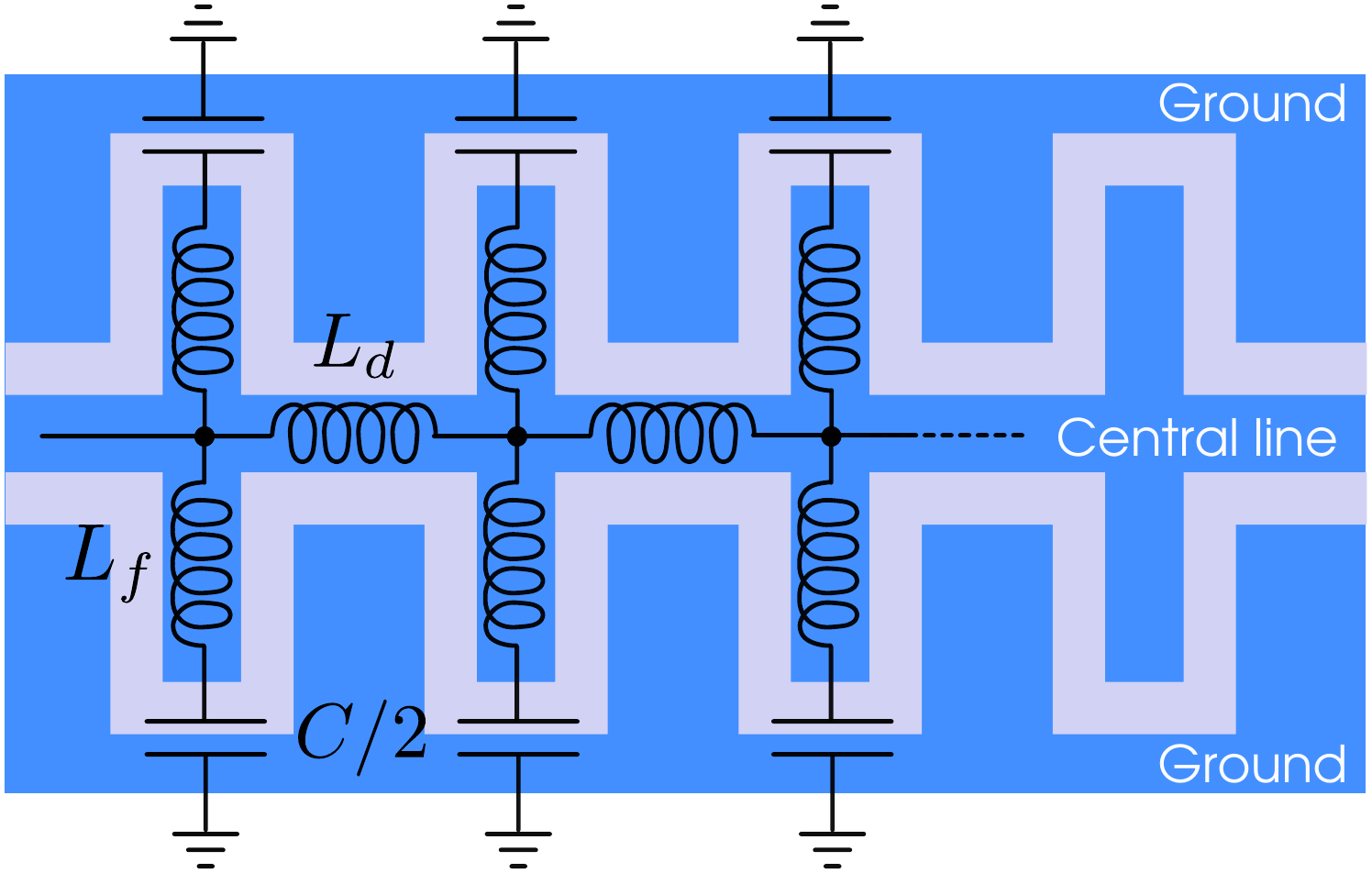}
\caption{\label{fig:KIT} Circuital model of a unit cell of a superconducting artificial line made of CPW sections (not to scale). In the equivalent electrical circuit, each cell is composed by a series inductance $L_d$ and two resonators with inductance $L_f$ and capacitance to ground $C/2$.}
\end{SCfigure}

Traveling wave amplifiers require both momentum conservation, i.e., phase matching, and energy conservation for the pump, signal, and any idlers that are generated. For the KI-TWPA device, momentum conservation can be attained by i) dispersion engineering of the CPW with periodic loadings that create a frequency gap or by ii) an artificial transmission line, also known as a lumped-element transmission line, that uses lumped-element inductors and capacitors instead of the distributed inductances and capacitances in a conventional transmission line. The characteristic impedance of the transmission line is modified every one-sixth of a wavelength at a frequency slightly above the pump frequency $f_p$ to form a wide stopband at $3f_p$. In addition, every third loading is modified in length (longer or shorter relative to the first two) to create a narrow stopband near $f_p$. Placed just below this narrow stopband, the pump tone picks up additional phase shift to fulfill the phase-matching condition, while its third harmonic is suppressed by the $3f_p$ stopband. As a result, exponential gain can be achieved over a broad bandwidth of more than one-half the pump frequency. 
CPW amplifiers, made in NbTiN, achieved 15\,dB gain with a transmission line 2\,m long~\cite{Eom2012}. More recent versions based on lumped elements, and made of NbTiN, achieved comparable gains with a transmission line 33 cm long~\cite{Malnou2021}. The advantage of the lumped element approach is that a shorter transmission line results in a higher fabrication yield~\cite{Chaudhuri2017}. In addition, the long CPW line could cause an impedance mismatch which is the likely cause of large ripples in the gain profile. As a result, the amplifier chip heats up, due to the strong pump tone, creating an excess of thermal noise. Considering all of these advantages, the goal of the DARTWARS project is to design KI-TWPA prototypes as a weakly dispersive artificial transmission line implemented by lumped-element inductors and capacitors (see Fig. \ref{fig:KIT} as example).

\section{KI-TWPA Fabrication}
The first aim is to tune the fabrication process in order to realize parametric amplifiers in NbTiN with a critical temperature around $T_c = 15$\,K~\cite{Chaudhuri2017} and a kinetic inductance around $L_k=10 $\,pH/square~\cite{Malnou2021}. To achieve these requirements a new sputtering system based on a KS 800 C Cluster provided by Kenosistec S.r.l. is currently being commissioned and tuned at the Fondazione Bruno Kessler - Center for Materials and Microsystems (FBK–CMM). The cluster is composed of two deposition chambers and a chamber for loading and transfers. Each deposition chamber is equipped with 4 Magnetron cathodes and turbomolecular pump able to reach a base vacuum of $5 \cdot 10^{-9}$\,mbar. The system is capable of handling 6" wafers and allows heating of substrates up to 400\,$^\circ$C. Deposition can be performed by reactive sputtering, by injecting Nitrogen, and by co-sputtering using two cathodes at the same time. 
In preliminary phases the system has been tested with deposition of Niobium (Nb) and Niobium nitride (NbN), finding a good match with the literature~\cite{Pinto2018}.

\begin{SCfigure}[][!t]
\includegraphics[clip=true,width=0.5\textwidth]{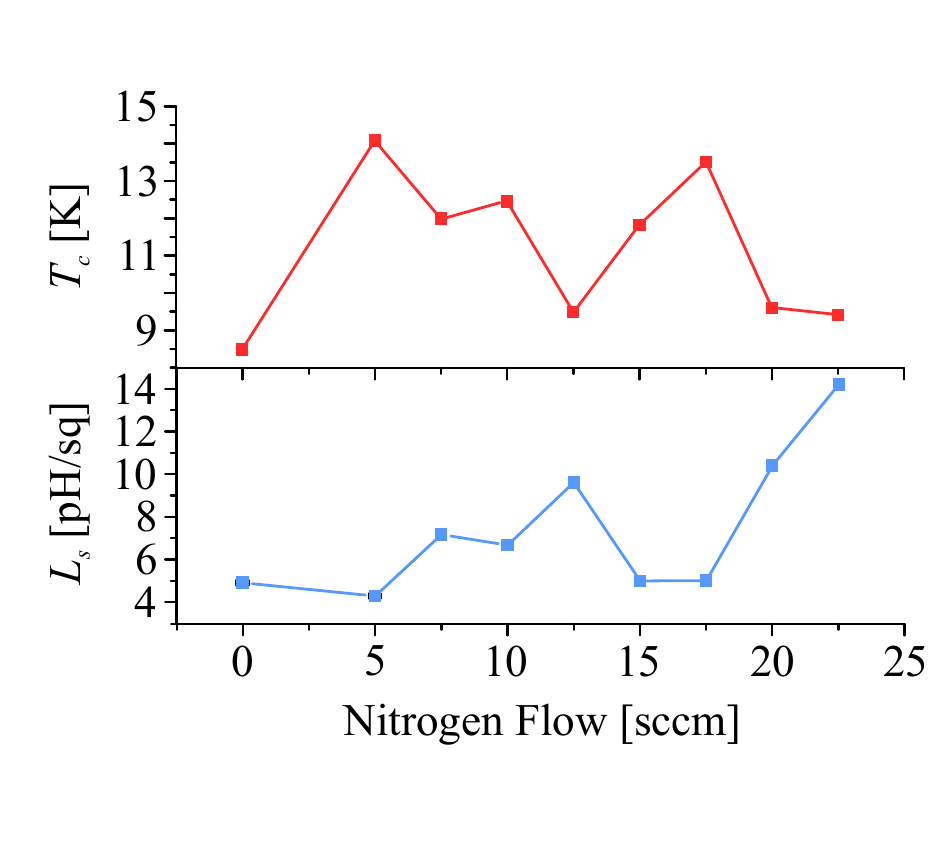}
\caption{\label{fig:Tc} (Top) Critical temperature as function of the nitrogen flow measured at cryogenic temperature for the deposited NbTiN films. (Bottom) Kinetic Inductance calculated based on the gap of the BCS theory and the sheet resistance and extrapolated for a film thickness of 30\,nm}
\end{SCfigure}

Most of the NbTiN film depositions have been done with the reactive sputtering technique using a target of Nb$_{0.66}$Ti$_{0.34}$ alloy which reduces the number of parameters to optimize. 
All films have been deposited on oxidized Si-wafers heated up to 400\,$^\circ$C. In these tests the chamber pressure was kept at $3 \cdot 10^{-3}$\,mbar with an Ar flow of 50\,sccm while the N$_2$ flow was verified from 5 to 22.5\,sccm. As shown in Fig. \ref{fig:Tc} the highest critical temperature of $T_c=14.08$\,K~\ has been obtained at a low N$_2$ flow with a very narrow transition of 4\,mK. This film resulted tensile with an estimated sheet resistance of 1312\,n\textOmega m. The residual resistance ratio of the films resulted in the order of 1. For all films the kinetic inductance has been calculated based on the gap of the BCS theory and the sheet resistance and extrapolated for a film thickness of 30\,nm , see Fig. \ref{fig:Tc}. The next tests will explore in more detail the region with the highest transition temperature and to see if it is possible to obtain more compressive films by rising the process pressure. The first KI-TWPA prototypes are foreseen in 2022 and will be produced exploiting the optimized deposition and etching procedures developed during 2021.


The first KI-TWPA prototypes are foreseen in 2022 and will be produced exploiting the optimized deposition and etching procedures developed during 2021.

\vspace{-1em}

\section{Conclusion and Future Plans}
TWPA are promising candidates as quantum limited microwave amplifiers for applications in fundamental physics experiments and quantum computing. The DARTWARS project aims to develop high performances TWPA by exploring new design solutions, new materials, and advanced fabrication processes. 

\begin{acknowledgements}
This work is supported by the European Union’s H2020-MSCA Grant Agreement No. 101027746, by the Italian Institute of Nuclear Physics (INFN) within the Technological and Interdisciplinary research commission (CSN5), by the Institute for Basic Science (IBS-R017-D1) of the Republic of Korea. and  by the Joint  Research  Project  PARAWAVE  of  the  European Metrology Programme for Innovation and Research (EMPIR). 

\end{acknowledgements}



\begin{thebibliography}{99}

\bibitem{Krantz2019}
P.~Krantz {\it et al.}, {\it Appl. Phys. Rev.} \textbf{6}, 021318 (2019),
DOI:\href{https://doi.org/10.1063/1.5089550}{10.1063/1.5089550} 

\bibitem{Stehlik2015}
J.~Stehlik {\it et al.}, {\it Phys. Rev. Applied.} \textbf{4}, 014018, (2015),
DOI:\href{https://doi.org/10.1103/PhysRevApplied.4.014018}{10.1103/PhysRevApplied.4.014018} 

\bibitem{Irwin-Hilton2005}
K. D. Irwin and G. C. Hilton,  {\it Topics in Applied Physics} \textbf{99} 63-149 (2005);
DOI:\href{https://doi.org/10.1007/10933596_3}{10.1007/10933596\_3} 

\bibitem{Gottardi2021}
L~Gottardi and K~Nagayashi, {\it Appl. Sci} \textbf{11}(9), 3793, (2021),
DOI:\href{https://doi.org/10.3390/app11093793}{10.3390/app11093793} 

\bibitem{Day2003}
P.~Day {\it et al.}, {\it Nature} \textbf{425} 817–821 (2003), 
DOI:\href{https://doi.org/10.1038/nature02037}{10.1038/nature02037}

\bibitem{Zmuidzinas2012}
J.~Zmuidzinas, {\it Annu. Rev. Condens. Matter Phys.} \textbf{3}(1), 169-214 (2012)
DOI:\href{https://doi.org/10.1146/annurev-conmatphys-020911-125022}{10.1146/annurev-conmatphys-020911-125022}

\bibitem{Alesini2019}
D.~Alesini {\it et al.}, {\it Phys. Rev. D} \textbf{99}(10), 101101 (2019),
DOI:\href{https://doi.org/10.1103/PhysRevD.99.101101}{10.1103/PhysRevD.99.101101}

\bibitem{Crescini2020}
N. Crescini {\it et al.}, {\it  Phys.Rev.Lett.} \textbf{124}(17), 171801, (2020),
DOI:\href{https://doi.org/10.1103/PhysRevLett.124.171801}{10.1103/PhysRevLett.124.171801}

\bibitem{Caves1982}
C~M.~Caves, {\it  Phys. Rev. D} \textbf{26}, 1817, (1982),
DOI:\href{https://doi.org/10.1103/PhysRevD.26.1817}{10.1103/PhysRevD.26.1817}


\bibitem{Castellanos2007}
M.~A.~Castellanos-Beltran and K.~W.~Lehnert, {\it Appl. Phys. Lett.} \textbf{91}, 083509 (2007),
DOI:\href{https://doi.org/10.1063/1.2773988}{10.1063/1.2773988}

\bibitem{Brien2014}
K. O'Brien {\it et al.}, {\it Phys. Rev. Lett.} \textbf{113}, 157001, (2014),
DOI:\href{https://doi.org/10.1103/PhysRevLett.113.157001}{10.1103/PhysRevLett.113.157001}

\bibitem{White2015}
T. C. White {\it et al.}, {\it Appl. Phys. Lett.} \textbf{106}, 242601, (2015),
DOI:\href{https://doi.org/10.1063/1.4922348}{10.1063/1.4922348}

\bibitem{Macklin2015}
C.~Macklin {\it et al.}, {\it Science} \textbf{350}(6258), 307-310 (2015),
DOI:\href{https://doi.org/10.1126/science.aaa8525}{10.1126/science.aaa8525}

\bibitem{Eom2012}
H.~Eom {\it et al.}, {\it Nature Phys.} \textbf{8}, 623–627, (2012),
DOI:\href{https://doi.org/10.1038/nphys2356}{10.1038/nphys2356}

\bibitem{Vissers2016}
M.~R.~Vissers {\it et al.}, {\it Appl. Phys. Lett.} \textbf{108}, 012601, (2016),
DOI:\href{https://doi.org/10.1063/1.4937922}{10.1063/1.4937922}

\bibitem{Zobrist2019}
N.~Zobrist {\it et al.}, {\it Appl. Phys. Lett.} \textbf{115}, 042601, (2019),
DOI:\href{https://doi.org/10.1063/1.5098469}{10.1063/1.5098469}

\bibitem{Malnou2021}
M.~Malnou {\it et al.}, {\it  PRX Quantum} \textbf{2}, 010302, (2021),
DOI:\href{https://doi.org/10.1103/PRXQuantum.2.010302}{10.1103/PRXQuantum.2.010302}

\bibitem{Zorin2016}
A.B.~Zorin, {\it Phys. Rev. Appl.} \textbf{6}, 034006 (2016),
DOI:\href{https://doi.org/10.1103/PhysRevApplied.6.034006}{10.1103/PhysRevApplied.6.034006}

\bibitem{Greco2020}
A.~Greco {\it et al.}, 
\href{https://arxiv.org/abs/2009.01002}{arXiv:2009.01002 [cond-mat.supr-con]}

\bibitem{Fasolo2021}
AL.~Fasolo {\it et al.}, 
\href{https://arxiv.org/abs/2109.14924}{arXiv:2109.14924 [cond-mat.supr-con]}

\bibitem{OBrien2014}
K.~O’Brien {\it et al.}, {\it Phys. Rev. Lett.} \textbf{113}, 157001 (2014),
DOI:\href{https://doi.org/10.1103/PhysRevLett.113.157001}{10.1103/PhysRevLett.113.157001}

\bibitem{Zorin2021}
A. B. Zorin, {\it Appl. Phys. Lett.} \textbf{118}, 222601 (2021),
DOI:\href{https://doi.org/10.1063/5.0050787}{10.1063/5.0050787}

\bibitem{Dixon2020}
T.~Dixon {\it et al.}, {\it Phys. Rev. Applied} \textbf{14}, 034058 (2020),
DOI:\href{https://doi.org/10.1103/PhysRevApplied.14.034058}{10.1103/PhysRevApplied.14.034058}

\bibitem{Chaudhuri2017}
S.~Chaudhuri {\it et al.}, {\it Appl. Phys. Lett.} \textbf{110}, 152601 (2017),
DOI:\href{https://doi.org/10.1063/1.4980102}{10.1063/1.4980102}

\bibitem{Pinto2018}
N. Pinto {\it et al.}, {\it Scientific Reports} \textbf{8}, 4710 (2018),
DOI:\href{https://doi.org/10.1038/s41598-018-22983-6}{10.1038/s41598-018-22983-6}

\end{thebibliography}
\end{document}